\documentclass[Physsubmission, Phys]{SciPost}
\hypersetup{
    colorlinks,
    linkcolor={red!50!black},
    citecolor={blue!50!black},
    urlcolor={blue!80!black}
}

\usepackage[bitstream-charter]{mathdesign}
\usepackage{graphicx}

\DeclareSymbolFont{usualmathcal}{OMS}{cmsy}{m}{n}
\DeclareSymbolFontAlphabet{\mathcal}{usualmathcal}

\begin{document}

% TODO: write your article's title here.
% The article title is centered, Large boldface, and should fit in two lines
\begin{center}{\Large \textbf{
Mergers as a Probe of Particle Dark Matter
}}\end{center}

% TODO: write the author list here. Use initials + surname format.
% Separate subsequent authors by a comma, omit comma at the end of the list.
% Mark the corresponding author with a superscript *.
\begin{center}
Anupam Ray\textsuperscript{1$\star$}
\end{center}

% TODO: write all affiliations here.
% Format: institute, city, country
\begin{center}
{\bf 1} Tata Institute of Fundamental Research, Homi Bhabha Road, Mumbai 400005, India
\\
% TODO: provide email address of corresponding author
* anupam.ray@theory.tifr.res.in
\end{center}

\begin{center}
\today
\end{center}

% For convenience during refereeing (optional),
% you can turn on line numbers by uncommenting the next line:
%\linenumbers
% You should run LaTeX twice in order for the line numbers to appear.

\definecolor{palegray}{gray}{0.95}
\begin{center}
\colorbox{palegray}{
  \begin{tabular}{rr}
  \begin{minipage}{0.1\textwidth}
    \includegraphics[width=30mm]{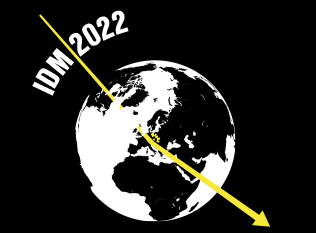}
  \end{minipage}
  &
  \begin{minipage}{0.85\textwidth}
    \begin{center}
    {\it 14th International Conference on Identification of Dark Matter}\\
    {\it Vienna, Austria, 18-22 July 2022} \\
    \doi{10.21468/SciPostPhysProc.12.056}\\
    \end{center}
  \end{minipage}
\end{tabular}
}
\end{center}

\section*{Abstract}
{\bf
% TODO: write your abstract here.
Black holes below Chandrasekhar mass limit (1.4 $M_{\odot}$) can not be produced via any standard stellar evolution. Recently, gravitational wave experiments have also discovered unusually low mass black holes whose origin is yet to be known. We propose a simple yet novel formation mechanism of such low mass black holes. Non-annihilating particle dark matter, owing to their interaction with stellar nuclei, can gradually accumulate inside compact stars, and eventually swallows them to low mass black holes, ordinarily impermissible by the Chandrasekhar limit. We point out several avenues to test this proposal, concentrating on the cosmic evolution of the binary merger rates.
}

% TODO: include a table of contents (optional)
% Guideline: if your paper is longer that 6 pages, include a TOC
% To remove the TOC, simply cut the following block
%\vspace{10pt}
\noindent\rule{\textwidth}{1pt}
\tableofcontents\thispagestyle{fancy}
\noindent\rule{\textwidth}{1pt}
%\vspace{10pt}
\section{Introduction}
\label{sec:intro}
% TODO: write your article here.
The recent observations of unusually low mass compact objects by the LIGO-VIRGO collaboration~\cite{LIGOScientific:2020aai,LIGOScientific:2020zkf,LIGOScientific:2021qlt} have ignited interest in the study of low mass black holes (BHs). More interestingly, standard stellar evolution cannot lead to a sub-Chandrasekhar mass BH, and the observation of such a BH  would herald new
physics. With immense improvement in gravitational
wave (GW) astronomy in recent times, the detection of a sub-Chandrasekhar mass BH is possibly forthcoming. Therefore, the key question, assuming a future GW observation involving a sub-Chandrasekhar mass BH, is how to identify its origin?
\\Primordial black holes (PBHs), with no compelling formation mechanisms,  are the most accepted explanation
of these objects~\cite{1967SvA....10..602Z,Hawking:1971ei,Chapline:1975ojl}. The existing alternative proposals, such as, accretion of fermionic asymmetric DM with non-negligible
self-interaction into compact stars~\cite{Kouvaris:2018wnh} or dark atomic cooling~\cite{Shandera:2018xkn} are not generic, and appeal to fairly convoluted
DM models. Transit of tiny PBHs (PBHs in the mass range of $10^{-15}-10^{-9} M_{\odot}$) through a compact star, and subsequent conversion of the compact star to a BH is also thought to be a novel mechanism to produce such low mass BHs~\cite{Capela:2013yf,Takhistov:2017bpt}. However, several recent works~\cite{Montero-Camacho:2019jte,Genolini:2020ejw} have falsified this proposal.\\ We point out a simple yet novel mechanism that
transmutes a sub-Chandrasekhar or $\mathcal{O} (1) M_{\odot}$ compact star to a comparable “low mass BH”. Non-annihilating particle DM  with non-zero interaction strength with the stellar nuclei, a universal feature of the DM models, is sufficient to produce  such low mass non-primordial BHs.  In the following, we briefly describe the formation mechanism of such low mass BHs, and answer a few basic questions, such as, what particle physics parameter space can they probe, how to test their origin?
\begin{figure}
	\centering
	\includegraphics[width=0.8\textwidth]{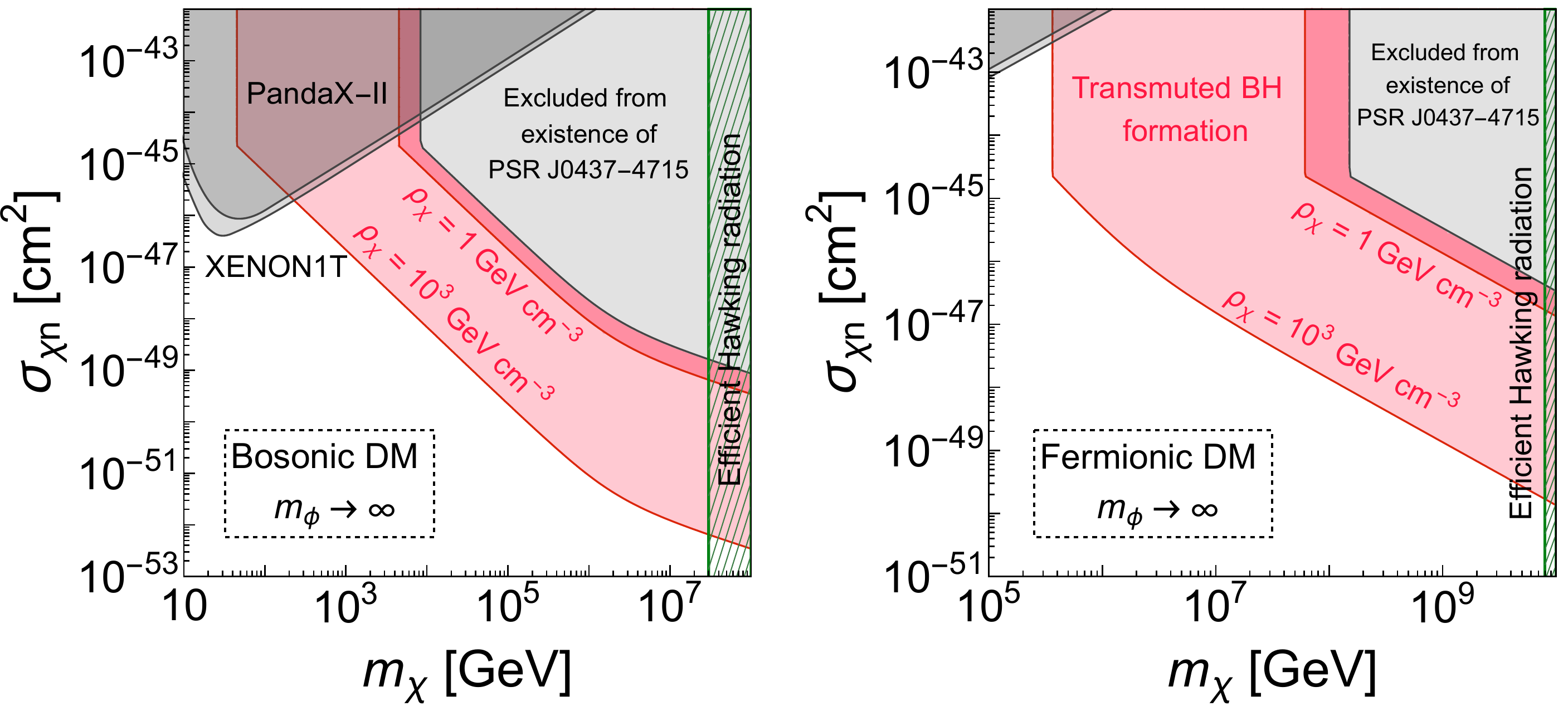}
	\caption{DM mass and scattering cross-section required for a dark core collapse and subsequent transmutation of a 1.3\,$M_{\odot}$ neutron star (NS) to a comparable mass BH are shown in the red shaded regions. The left (right) panel is for bosonic (fermionic) DM, and a contact interaction between DM and the stellar nuclei is assumed. Two representative values of ambient DM density, $\rho_{\chi}$\,=\,1~and~$10^3$\,GeV\,cm$^{-3}$ are considered.  Exclusion limits from the underground direct detection experiments { PandaX-II}\,\cite{PandaX-II:2018xpz} and { XENON1T}\,\cite{XENON:2018voc} as well as from existence of an $\sim$ 7 Gyr old~\cite{Manchester:2004bp} nearby pulsar PSR\,J0437-4715\,\cite{Dasgupta:2020dik,McDermott:2011jp,Garani:2018kkd} are also shown by the gray shaded regions. Green hatched regions denote the parameter space where efficient Hawking evaporation ceases the implosion of the NS. The figure is taken from~\cite{Dasgupta:2020mqg}.}
	\label{fig: particle physics constraints}
\end{figure}
\section{Formation of Low Mass Transmuted Black Holes}
Non-annihilating particle dark matter (DM)~\cite{Petraki:2013wwa,Zurek:2013wia}, owing to their interactions with the stellar nuclei, can accumulate inside stellar objects via single~\cite{Press:1985ug,gould1,Bell:2020lmm,Bell:2020obw} or multiple~\cite{Bramante:2017xlb,Ilie:2020vec,Dasgupta:2019juq} scatterings with the stellar targets. Inside the stellar core, the captured number of DM particles grow linearly with time. Once the total number of captured DM particles $\left(N_{\chi} \rvert_{ t_{\rm{age}}}\right)$ throughout the age of the stellar object $\left(t_{\rm{age}}\right)$ satisfies the BH formation criterion, i.e., $N_{\chi} \rvert_{ t_{\rm{age}}} \geq \max \left[ N^{\rm{self}}_{\chi}, N^{\rm{cha}}_{\chi}\right]$, it ensues a dark core collapse, eventually transmuting the hosts to comparable mass BHs. $N^{\text{self}}_{\chi}$ denotes  the required number of DM particles for self-gravitation, and is set by the condition that the captured DM density within the stellar core has to exceed the corresponding baryonic density~\cite{McDermott:2011jp}. Whereas, Chandrasekhar limit, $N^{\rm cha}_{\chi}$, depends on the spin-statistics of the DM particles, and it is much easier to achieve for bosonic DM as compared to fermionic DM, explaining an easier transmutation for  bosonic DM.\\ 
Once the number of captured DM particles satisfies the BH formation criterion, dark core collapse initiates, and a tiny BH forms inside the stellar object. This tiny BH accumulates matter from the host, and eventually swallows the host to a comparable mass BH in a very short timescale~\cite{Baumgarte:2021thx,Richards:2021upu,Schnauck:2021hlm,Giffin:2021kgb}. Such BHs are  known as  transmuted black holes (TBHs), and depending on the mass of the progenitors, TBHs can naturally be sub-Chandrasekhar, or even sub-solar.  However, note that, if the nascent BH that forms via dark core collapse is sufficiently light, it quickly evaporates due to its efficient Hawking emission, ceasing the transmutation. For typical neutron star (NS) parameters, if the initial BH mass is lighter than $\sim 10^{-20}\,{M}_{\odot}$, Hawking evaporation dominates over the swallowing process, and the transmutation ceases~\cite{Dasgupta:2020dik, Kouvaris:2010jy}. For non-annihilating bosonic and fermionic DM, it corresponds to DM masses $\gtrsim \mathcal O (10^7)$ and $\gtrsim \mathcal O (10^{10})$ GeV, respectively, providing an upper limit on the DM mass for transmutation. In Fig.\,\ref{fig: particle physics constraints} we demonstrate the DM parameter space where a NS with mass $1.3\,{M}_{\odot}$ can transmute to a low mass BH for either bosonic or fermionic DM, for two choices of ambient DM density.
\begin{figure}[!t]
	\includegraphics[width=0.5\textwidth]{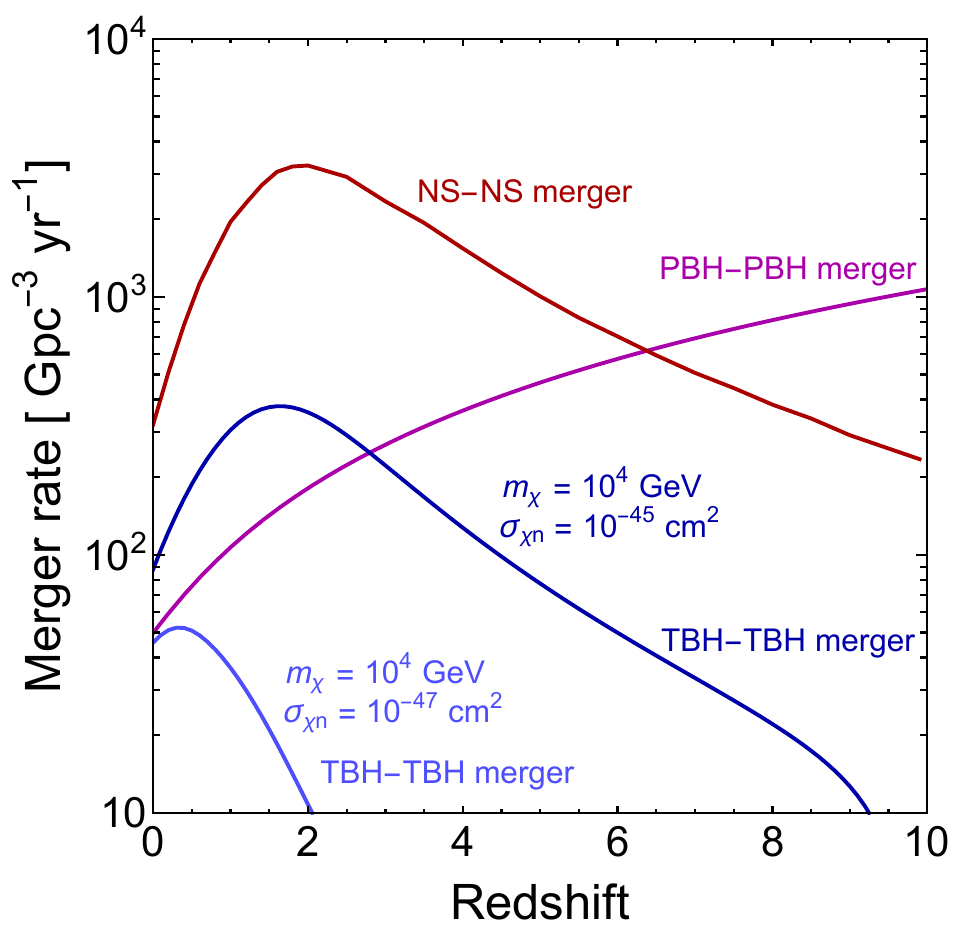}
	\centering
	\caption{Cosmic evolution of the binary merger rates provides a novel technique to determine the stellar or primordial origin of low mass BHs. Cosmic evolution of the binary PBH, NS, and TBH merger rates are shown in the redshift range of 0 to 10. For the binary NS and TBH merger rate, cosmic star formation rate is adopted from~\cite{Madau:2014bja} and they are normalized to the recent LIGO-VIRGO measurement~\cite{LIGOScientific:2020kqk}. Non-annihilating bosonic DM with mass of 10 TeV and DM-nucleon scattering cross-section of $10^{-45}$ and $10^{-47}$ $\textrm{cm}^2$ in the contact approximation are assumed for the estimation of binary TBH merger rate. The PBH merger rate is estimated by considering 1.3 $M_{\odot} -$ 1.3 $M_{\odot}$ PBH binary and a DM fraction $f_{\rm PBH}=10^{-3}$. The figure is taken from~\cite{Dasgupta:2020mqg}.}
	\label{fig:merger rate}
\end{figure} 
\section{Identifying the Origin of Low Mass Black Holes}
Formation of sub-Chandrasekhar mass non-primordial BHs via gradual accumulation of non annihilating DM in compact stars demand a critical investigation to pinpoint the origin of the low mass BHs. In this following, we briefly describe how cosmic evolution of the binary merger rates can be used to determine the origin of low mass BHs.
The merger rate of PBH binaries keeps rising with higher redshift, and it has a universal time dependence of $R_{\rm PBH} \propto t^{-34/37}$, where $t$ is the coalescence time at formation~\cite{Chen:2018czv,Raidal:2018bbj,Sasaki:2016jop,Ali-Haimoud:2017rtz,Sasaki:2018dmp}.  On the other hand, the merger rate of binary NSs, $R_{\rm{NS}} (t)$\,\cite{Taylor:2012db}, as shown in Fig.~\ref{fig:merger rate}, follows the cosmic star formation rate\,\cite{Madau:2014bja,Porciani:2000ag}
\begin{align}
	R_{\rm{NS}} (t) = \int_{t_f=t_*}^{t} dt_f \frac{dP_m}{dt}(t-t_f)\lambda\frac{d\rho_*}{dt}(t_f)\,.
	\label{eq:NS}
\end{align}
It peaks at an $\mathcal O$(1) redshift when the star formation rate is maximal. In Eq.\,(\ref{eq:NS}), $\lambda = 10^{-5} M^{-1}_{\odot}$ is the number of merging NS binaries per unit star forming mass, $\frac{d\rho_*}{dt} (t_f)$ denotes the cosmic star formation rate at the binary formation time $t_f$~\cite{Madau:2014bja},  and $\frac{dP_m}{dt} (t-t_f) \propto (t-t_f)^{-1}$ denotes the probability density distribution of coalescing BNSs within the time interval $(t-t_f)$ after formation. The earliest star formation time $t_*$ is taken as $4.9 \times 10^8$ year which corresponds to $z_*=10$~\cite{Taylor:2012db}.\\
The merger rate of TBH binaries, $R_{\rm{TBH}} (t)$, depends on the particle DM parameters such as DM mass $(m_{\chi})$, and DM-nucleon interaction strength $(\sigma_{\chi n})$ via the transmutation time $(\tau_{\rm{trans}})$, as well as on the astrophysical parameters such as the merger rate of binary NSs. 
$R_{\rm{TBH}} (t)$ is systematically lower than  $R_{\rm{NS}} (t)$, as only a fraction of the binary NS implode depending on the time required for transmutation. This fraction depends on the binary NS population in the galaxies, as well as evolution of the DM density in the galaxies, and it gradually decreases with higher redshifts as
NS binaries at higher redshift do not have the sufficient time to accumulate enough DM
required for implosion. Hence, $R_{\rm{TBH}} (t) $ takes the form
\begin{equation}
	R_{\rm{TBH}} (t) = \sum_{i} f_i \int_{t_f=t_*}^{t} dt_f \frac{dP_m}{dt}(t-t_f) \lambda\frac{d\rho_*}{dt}(t_f) \times  \Theta\left\{t-t_f-\tau_{\rm{trans}}\left [m_\chi,\sigma_{\chi n},\rho_{\textrm{ext},i}(t)\right] \right\}\,,
	\label{eq:TBH}
\end{equation}
In Eq.\,(\ref{eq:TBH}), we assume that the binary NSs reside in Milky-Way-like galaxies, and are uniformly distributed in $r=(0.01,0.1)$ kpc, where $r$ is the Galactocentric distance. We also assume that the DM density in each halo (at all redshifts) follows the Navarro-Frenk-White profile~\cite{Navarro:1995iw,Navarro:1996gj}, and the parameters of the Navarro-Frenk-White profile is essentially determined by the time evolution of the Hubble parameter.
From the expression for the merger rate, it is evident that  $R_{\rm{TBH}} (t)$ decreases with increase in transmutation time. Therefore, for a given a DM mass, decrease in DM-nucleon scattering cross-section leads to higher $\tau_{\rm{trans}}$, and, hence, lower $R_{\rm{TBH}}$, as shown in Fig.~\ref{fig:merger rate}. This distinct redshift dependence of the binary merger rates, particularly at higher redshifts, can be measured with the imminent ground as well as space-based GW detectors like {Cosmic Explorer}~\cite{Evans:2021gyd}, {Einstein Telescope}\,(ET)~\cite{Maggiore:2019uih}, and {Pre-DECIGO}\,\cite{Nakamura:2016hna}, enabling them to distinguish the transmutation scenario from PBHs. 
\section{Conclusion}
Sub-Chandrasekhar mass BHs cannot be described via any standard stellar evolution and will augur new physics. The existing alternative proposals are either not effective or appeal to fairly convoluted DM models. Here, we study a simple yet novel production mechanism for sub-Chandrasekhar mass non-primordial BHs. Gradual accumulation of non-annihilating particle DM inside compact stars can lead to transmutation of compact stars via dark core collapse, and that can give rise to low mass BHs. Cosmic evolution of the binary merger rates can be used as a novel probe to determine the origin of such low mass BHs. We demonstrate that measurement of the high-redshift binary merger rates by the imminent GW detectors can conclusively shed light on this topic.
\section*{Acknowledgments}
A.R. wishes to thank his collaborators Basudeb Dasgupta, Aritra Gupta, and  Ranjan Laha for valuable contributions in the original works~\cite{Dasgupta:2019juq,Dasgupta:2020dik,Dasgupta:2020mqg}.
\bibliography{ref.bib}
\end{document}